\magnification=1200

\hsize=125mm
\vsize=180mm
\hoffset=3mm
\voffset=12mm

\baselineskip=14pt

\font\twelvebf=cmbx12

\font\tenbb=msym10
\font\sevenbb=msym7
\font\fivebb=msym5
\newfam\bbfam
\textfont\bbfam=\tenbb \scriptfont\bbfam=\sevenbb
\scriptscriptfont\bbfam=\fivebb
\def\bb{\fam\bbfam}

\def\Cb{{\bb C}}
\def\Hb{{\bb H}}
\def\Rb{{\bb R}}
\def\Ab{{\bb A}}
\def\Bb{{\bb B}}
\def\Eb{{\bb E}}
\def\Zb{{\bb Z}}

\def\Ac{{\cal A}}
\def\Hc{{\cal H}}
\def\Lc{{\cal L}}
\def\Uc{{\cal U}}

\def\Diff{\mathop{\rm Diff}\nolimits}
\def\Aut{\mathop{\rm Aut}\nolimits}
\def\Trace{\mathop{\rm Trace}\nolimits}

\def\Tr{\mathop{\rm Tr}\nolimits}
\def\Re{\mathop{\rm Re}\nolimits}
\def\Res{\mathop{\rm Res}\nolimits}
\def\Gev{\mathop{\rm Gev}\nolimits}
\def\range{\mathop{\rm range}\nolimits}
\def\Int{\mathop{\rm Int}\nolimits}
\def\Out{\mathop{\rm Out}\nolimits}
\def\Sup{\mathop{\rm Sup}\nolimits}

\def\Lb{\Lambda}
\def\lb{\lambda}
\def\g{\gamma}
\def\vp{\varphi}
\def\a{\alpha}
\def\b{\beta}
\def\g{\gamma}
\def\G{\Gamma}
\def\s{\sigma}
\def\Si{\Sigma}
\def\om{\omega}
\def\Om{\Omega}
\def\d{\delta}
\def\t{\theta}
\def\k{\kappa}
\def\ve{\varepsilon}

\def\ts{\times}
\def\ify{\infty}
\def\fl{\forall}
\def\op{\oplus}
\def\ra{\rightarrow}
\def\wt{\widetilde}
\def\ot{\otimes}
\def\part{\partial}
\def\lgl{\langle}
\def\rgl{\rangle}
\def\nb{\nabla}
\def\eqv{\equiv}

\def\sm{\simeq}
\def\sbs{\subset}

\def\un{{\rm 1\mkern-4mu I}}

\def\up#1{\raise 1ex\hbox{\sevenrm#1}}

\def\semi{\mathop{>\!\!\!\triangleleft}}

\catcode`\@=11
\def\displaylinesno #1{\displ@y\halign{
\hbox to\displaywidth{$\@lign\hfil\displaystyle##\hfil$}&
\llap{$##$}\crcr#1\crcr}}

\vglue 1cm

\centerline{\twelvebf The Spectral Action Principle}

\vglue 1cm

\centerline{Ali H. Chamseddine\up{1,2} \quad and \quad
Alain Connes\up{2}}

\vglue 1cm

\noindent { 1. Theoretische Physik,
ETH-H\"onggerberg, CH-8093 Z\"urich, Switzerland}

\noindent {     2. I.H.E.S., F-91440
Bures-sur-Yvette, France}

\vglue 3cm

\noindent {\parindent=1cm\narrower
{\centerline {\bf Abstract.}} We propose a new action
principle to be associated with a noncommutative space
$(\Ac ,\Hc ,D)$. The universal formula for the spectral
action is $(\psi ,D\psi) + \Trace (\chi (D /$ $\Lb))$
where $\psi$ is a spinor on the Hilbert space, $\Lb$ is a
scale and $\chi$ a positive function. When this principle is
applied to the noncommutative space defined by the
spectrum of the standard model one obtains the standard
model action coupled to Einstein plus Weyl gravity. There
are relations between the gauge coupling constants
identical to those of $SU(5)$ as well as the Higgs
self-coupling, to be taken at a fixed high energy scale.
}

\vfill\eject

\noindent {\bf 1. Introduction.}

\medskip

The basic data of Riemannian geometry consists in a {\it
manifold} $M$ whose points $x\in M$ are locally labelled by
finitely many coordinates $x^{\mu} \in \Rb$, and in the
infinitesimal {\it line element}, $ds$,
$$
ds^2 = g_{\mu \nu} \, dx^{\mu} \, dx^{\nu} \, . \eqno (1.1)
$$
The laws of physics at reasonably low energies are well
encoded by the action functional,
$$
I = I_E + I_{SM} \eqno (1.2)
$$
where $I_E = {1 \over 16 \pi G} \int R \, \sqrt g \, d^4 x$
is the Einstein action, which depends only upon the
4-geometry (we shall work throughout in the Euclidean, i.e.
imaginary time formalism) and where $I_{SM}$ is the standard
model action, $I_{SM} = \int \Lc_{SM}$, $\Lc_{SM} = \Lc_G +
\Lc_{GH} + \Lc_H + \Lc_{Gf} + \Lc_{Hf}$. The action
functional $I_{SM}$ involves, besides the 4-geometry,
several additional fields: bosons $G$ of spin 1 such as $\g$,
$W^{\pm}$ and $Z$, and the eight gluons, bosons of spin 0
such as the Higgs field $H$ and fermions $f$ of spin $1/2$,
the quarks and leptons.

\smallskip

\noindent These additional fields have {\it a priori} a very
different status than the geometry $(M,g)$ and the gauge
invariance group which governs their interaction is {\it a
priori} very different from the diffeomorphism group which
governs the invariance of the Einstein action. In fact the
natural group of invariance of the functional (1.2) is the
semidirect product,
$$
G = \Uc \semi \Diff (M) \eqno (1.3)
$$
of the group of local gauge transformations, $\Uc = C^{\ify}
(M,U(1) \ts SU(2) \ts SU(3))$ by the natural action of
$\Diff (M)$.

\smallskip

\noindent The basic data of noncommutative geometry consists
of an involutive {\it algebra} $\Ac$ of operators in Hilbert
space $\Hc$ and of a selfadjoint unbounded operator $D$ in
$\Hc$ [1-6].

\smallskip

\noindent The inverse $D^{-1}$ of $D$ plays the role of the
infinitesimal unit of length $ds$ of ordinary geometry.

\smallskip

\noindent To a Riemannian compact spin manifold corresponds
the {\it spectral triple} given by the algebra $\Ac =
C^{\ify} (M)$ of smooth functions on $M$, the Hilbert space
$\Hc = L^2 (M,S)$ of $L^2$-spinors and the Dirac operator
$D$ of the Levi-Civita Spin connection. The line element
$ds$ is by construction the propagator of fermions,
$$
ds = \ts \!\! \hbox{---} \!\!\! \ts \, .  \eqno (1.4)
$$
No information is lost in trading the original Riemannian
manifold $M$ for the corresponding spectral triple $(\Ac
,\Hc ,D)$. The points of $M$ are recovered as the characters
of the involutive algebra $\Ac$, i.e. as the homomorphisms
$\rho : \Ac \ra \Cb$ (linear maps such that $\rho (ab) =
\rho (a) \, \rho (b) \quad \fl \, a,b \in \Ac$). The geodesic
distance between points is recovered by
$$
d(x,y) = \Sup \ \{ \vert a(x) - a(y)\vert \ ; \ a\in \Ac \ , \
\Vert [D,a]\Vert \leq 1 \} \, . \eqno (1.5)
$$
More importantly one can characterize the spectral triples
$(\Ac ,\Hc ,D)$ which come from the above spinorial
construction by very simple axioms ([4]) which involve the
dimension $n$ of $M$. The parity of $n$ implies a $\Zb /2$
grading $\g$ of the Hilbert space $\Hc$ such that,
$$
\g = \g^* \ , \ \g^2 =1 \ , \ \g a = a \g \quad \fl \, a \in
\Ac \ , \ \g D =-D \g \, . \eqno (1.6)
$$
Moreover one keeps track of the {\it real structure} on
$\Hc$ as an antilinear isometry $J$ in $\Hc$ satisfying
simple relations
$$
J^2 = \ve \ , \ JD = \ve' DJ \ , \ J\g = \ve'' \g J \ ; \
\ve ,\ve' ,\ve'' \in \{ -1,1\} \eqno (1.7)
$$
where the value of $\ve ,\ve' ,\ve''$ is determined by $n$
modulo 8. One first virtue of these axioms is to allow for a
shift of point of view, similar to Fourier transform, in
which the usual emphasis on the points $x\in M$ of a
geometric space is now replaced by the spectrum $\Si \sbs
\Rb$ of the operator $D$. Indeed, if one forgets about the
algebra $\Ac$ in the spectral triple $(\Ac ,\Hc ,D)$ but
retains only the operators $D$, $\g$ and $J$ acting in $\Hc$
one can (using (1.7)) characterize this data by the spectrum
$\Si$ of $D$ which is a discrete subset with multiplicity of
$\Rb$. In the even case $\Si = -\Si$. The existence of
Riemannian manifolds which are isospectral (i.e. have the
same $\Si$) but not isometric shows that the following
hypothesis is stronger than the usual diffeomorphism
invariance of the action of general relativity,
$$
\hbox{``The physical action only depends upon} \ \Si \,
.\hbox{''} \eqno (1.8)
$$
In order to apply this principle to the action (1.2) we need
to exploit a second virtue of the axioms (cf. [4]) which is
that they do not require the commutativity of the algebra
$\Ac$. Instead one only needs the much weaker form,
$$
ab^0 = b^0 a \quad \fl \, a,b \in \Ac \quad \hbox{with}
\quad b^0 = Jb^* J^{-1} \, . \eqno (1.9)
$$
In the usual Riemannian case the group $\Diff (M)$ of
diffeomorphisms of $M$ is canonically isomorphic to the group
$\Aut (\Ac)$ of automorphisms of the algebra $\Ac = C^{\ify}
(M)$. To each $\vp \in \Diff (M)$ one associates the algebra
preserving map $\a_{\vp} : \Ac \ra \Ac$ given by
$$
\a_{\vp} (f) = f \circ \vp^{-1} \qquad \fl \, f \in C^{\ify}
(M) = \Ac \, . \eqno (1.10)
$$
In general the group $\Aut (\Ac)$ of automorphisms of the
involutive algebra $\Ac$ plays the role of the
diffeomorphisms of the noncommutative (or spectral for
short) geometry $(\Ac ,\Hc ,D)$. The first interesting new
feature of the general case is that the group $\Aut (\Ac)$
has a natural normal subgroup,
$$
\Int (\Ac ) \sbs \Aut (\Ac) \eqno (1.11)
$$
where an automorphism $\a$ is {\it inner} iff there exists a
unitary operator $u\in \Ac$, $(uu^* = u^* u=1)$ such that,
$$
\a (a) = uau^* \qquad \fl \, a \in \Ac \, . \eqno (1.12)
$$
The corresponding exact sequence of groups,
$$
1 \ra \Int (\Ac) \ra \Aut (\Ac) \ra \Out (\Ac) \ra 1 \eqno
(1.13)
$$
looks very similar to the exact sequence
$$
1\ra \Uc \ra G \ra \Diff (M) \ra 1 \eqno (1.14)
$$
which describes the structure of the symmetry group $G$ of
the action functional (1.2).

\smallskip

\noindent Comparing (1.13) and (1.14) and taking into account the
action of inner automorphisms of $\Ac$ in $\Hc$ given by
$$
\xi \ra u(u^*)^0 \xi = u\xi u^* \eqno (1.15)
$$
one determines the algebra $\Ac$ such that $\wt{\Aut} (\Ac)
=G$ (where $\wt{\Aut}$ takes into account the action of
automorphisms in the Hilbert space $\Hc$). The answer is
$$
\Ac = C^{\ify} (M) \ot \Ac_F \eqno (1.16)
$$
where the algebra $\Ac_F$ is {\it finite dimensional},
$$
\Ac_F = \Cb \op \Hb \op M_3 (\Cb) \eqno (1.17)
$$
where $\Hb \sbs M_2 (\Cb)$ is the algebra of quaternions,
$$
\Hb = \left\{ \left( \matrix{\a &\b \cr -\bar{\b} &\bar{\a}
\cr }  \right) \ ; \ \a ,\b \in \Cb \right\} \, . \eqno (1.18)
$$
Giving the algebra $\Ac$ does not suffice to determine the
spectral geometry, one still needs the action of $\Ac$ in
$\Hc$ and the operator $D$. Since $\Ac$ is a tensor product
(16) which geometrically corresponds to a product space, an
instance of spectral geometry for $\Ac$ is given by the
product rule,
$$
\Hc = L^2 (M,S) \ot \Hc_F \ , \ D = {\part \!\!\! /}_M \ot 1
+ \g_5 \ot D_F \eqno (1.19)
$$
where $(\Hc_F ,D_F)$ is a spectral geometry on $\Ac_F$,
while both $L^2 (M,S)$ and the Dirac operator ${\part \!\!\!
/}_M$ on $M$ are as above.

\smallskip

\noindent Since $\Ac_F$ is finite dimensional the dimension
of the corresponding space is $0$ so that $\Hc_F$ must be
finite dimensional. The list of elementary fermions provides
a natural candidate for $\Hc_F$. One lets $\Hc_F$ be the
Hilbert space with basis labelled by elementary leptons and
quarks. Thus for the first generation of leptons we get
$e_L, e_R ,\nu_L ,\bar{e}_L ,\bar{e}_R ,\bar{\nu}_L$ for
instance, as the corresponding basis. The $\Zb /2$ grading
$\g_F$ is given by $+1$ for left handed particles and $-1$
for right handed ones. For quarks one has an additional
color index, $y,r,b$. The involution $J$ is just such that
$Jf = \bar f$ for any $f$ in the basis. One has $J^2 =1$,
$J\g = \g J$ as dictated by the dimension $n=0$. Moreover
the algebra $\Ac_F$ has a natural representation in $\Hc_F$
and:
$$
ab^0 = b^0 a \qquad \fl \, a,b \in \Ac_F \ , \ b^0 = Jb^*
J^{-1} \, . \eqno (1.20)
$$
Finally there is a natural matrix acting in the finite
dimensional Hilbert space $\Hc_F$. It is
$$
D_F = \left[ \matrix{
Y &0 \cr 0 &\bar Y \cr
} \right] \, , \eqno (1.21)
$$
where $Y$ is the Yukawa coupling matrix.

\smallskip

\noindent The special features of $Y$ show that the
algebraic rule
$$
[[D,a],b^0 ]=0 \qquad \fl \, a,b \in \Ac \eqno (1.22)
$$
which is one of the essential axioms, holds for the spectral
geometry $(\Ac_F ,\Hc_F ,D_F)$ $=F$. Of course this
0-dimensional geometry is encoding the knowledge of the
fermions of the standard model and it is a basic question to
understand and characterize it abstractly, but let us
postpone this problem and proceed with the product geometry
$M\ts F$.

\smallskip

\noindent The next important new feature of the
noncommutative case is the following. We saw that the group
$\Aut (\Ac)$ of diffeomorphisms falls in equivalence classes
under the normal subgroup $\Int (\Ac)$ of inner
automorphisms. In the same way the space of metrics has a
natural foliation into equivalence classes. The {\it
internal fluctuations} of a given metric are given by the
formula,
$$
D=D_0 + A + JAJ^{-1} \ , \ A = \Si \, a_i [D_0 ,b_i] \ , \ a_i
, b_i \in \Ac \ \hbox{and} \ A=A^* \, . \eqno (1.23)
$$
Thus starting from $(\Ac ,\Hc ,D_0)$ with obvious notations,
one leaves the representation of $\Ac$ in $\Hc$ untouched
and just perturbs the operator $D_0$ by (1.23) where $A$ is an
arbitrary self-adjoint operator in $\Hc$ of the form $A=\Si
\, a_i [D_0 ,b_i] \ ; \ a_i ,b_i \in \Ac$. One checks that
this does not alter the axioms (check (1.22) for instance).
These fluctuations are trivial: $D=D_0$ in the usual
Riemannian case in the same way as the group of inner
automorphisms $\Int (\Ac) = \{ 1\}$ is trivial for $\Ac =
C^{\ify} (M)$.

\smallskip

\noindent In general the natural action of $\wt{\Int} (\Ac)$
on the space of metrics restricts to the above equivalence
classes and is simply given by (for the automorphism
associated to $u\in \Ac$, $uu^* = u^* u =1$),
$$
\xi \in \Hc \ra u\xi u^* =uu^{*0} \, \xi \ , \ A \ra
u[D,u^*] + uAu^* \, . \eqno (1.24)
$$
When one computes the internal fluctuations of the above
product geometry $M\ts F$ one finds ([6]) that they are
parametrized exactly by the bosons $\g ,W^{\pm} ,Z$, the
eight gluons and the Higgs fields $H$ of the standard model.
The equality
$$
\int_M (\Lc_{Gf} + \Lc_{Hf}) \, \sqrt g \, d^4 x = \lgl \psi
,D\psi \rgl \eqno (1.25)
$$
gives the contribution to (1.2) of the last two term of the
$SM$ Lagrangian in terms of the operator $D$ alone.

\smallskip

\noindent The operator $D$ encodes the metric of our
``discrete Kaluza Klein'' geometry $M\ts F$ but this metric
is no longer the product metric as it was for $D_0$. In fact
the initial scale given by $D_F$ completely disappears when
one considers the arbitrary internal fluctuations of $D_0 =
{\part \!\!\! /}_M \ot 1 + \g_5 \ot D_F$. What remains is to
understand in a purely gravitational manner the 4 remaining
terms of the action (1.2). This is where we apply the basic
principle (1.8).

\smallskip

\noindent We shall check in this paper that for any smooth
cutoff function $\chi$, $\chi (\lb) =1$ for $\vert \lb \vert
\leq 1$, one has
$$
\Trace \chi \left( {D\over \Lb} \right) = I_E + I_G + I_{GH}
+ I_H + I_C + 0(\Lb^{-\ify}) \eqno (1.26)
$$
where $I_C$ is a sum of a cosmological term, a term of Weyl
gravity and a term in $\int R \, H^2 \, \sqrt g \, d^4 x$. The
computation in itself is not new, and goes back to the work
of DeWitt [7]. Similar computations also occur in the theory
of induced gravity [8]. It is clear that the left hand side
of (1.26) only depends upon the spectrum $\Si$ of the operator
$D$, and following our principle (1.8) this allows to take it
as the natural candidate for the bare action at the cutoff
scale $\Lb$.

\smallskip

\noindent In our framework there is a natural way to cutoff
the geometry at a given energy scale $\Lb$, which has been
developed in [9] for some concrete examples. It consists
in replacing the Hilbert space $\Hc$ by the subspace
$\Hc_{\Lb}$,
$$
\Hc_{\Lb} = \range \chi \left( {D\over \Lb} \right) \eqno
(1.27)
$$
and restricting both $D$ and $\Ac$ to this subspace, while
maintaining the commutation rule (1.20) for the algebra $\Ac$.
This procedure is superior to the familiar lattice
approximation because it does respect the geometric symmetry
group. The point is that finite dimensional {\it
noncommutative} algebras have continuous Lie groups of
automorphisms while the automorphism group of a commutative
finite dimensional algebra is necessarily finite. The
hypothesis which we shall test in this paper is that there
exist an energy scale $\Lb$ in the range $10^{15} -
10^{19} \Gev$ at which the bare action (1.2) becomes geometric,
i.e. $\sim$ 
$$
\Trace \chi \left( {D\over \Lb} \right) + \lgl \psi ,D\psi
\rgl \, . \eqno (1.28)
$$

\vglue 1cm

\noindent {\bf 2. The spectral action principle applied
to the Einstein-Yang-Mills system.}

\medskip

To test the spectral action functional (1.28) we
shall first consider the simplest noncommutative
modification of a  manifold $M$. Thus we replace the algebra
$C^{\ify} (M)$ of smooth functions on $M$ by the tensor
product $\Ac = C^{\ify} (M) \ot M_N (\Cb)$ where
$M_N (\Cb)$ is the algebra of $N\ts N$ matrices. The spectral triple is
obtained by tensoring the Dirac spectral triple for $M$ by
the spectral triple for $M_N (\Cb)$ given by the left action
of $M_N (\Cb)$ on the Hilbert space of $N\ts N$ matrices
with Hilbert-Schmidt norm while the operator is 0. The real
structure is given by the adjoint operation, $m\ra m^*$ on
matrices. Thus for the product geometry one has
$$
\displaylinesno
{
\Hc = L^2 (M,S) \ot M_N (\Cb) \cr
J(\xi \ot m) = C\xi \ot m^* &(2.1) \cr
D = {\part \!\!\! /}_M \ot 1 \, . \cr
}
$$
We shall compare the spectral action functional (1.28) with
the following
$$
I = {1\over 2\k^2} \int R \, \sqrt g \, d^4 x + I_{YM}
\eqno (2.2)
$$
where $I_{YM} = \int (\Lc_G + \Lc_{Gf}) \, \sqrt g \, d^4 x$
is the action for an $SU(N)$ Yang-Mills theory coupled to
fermions in the adjoint representation.

\smallskip

\noindent Starting with (2.1), one first computes the
internal fluctuations of the metric and finds that they are
parametrized exactly by an $SU(N)$ Yang Mills field $A$.
Note that the formula $D=D_0 +A+JAJ^*$ eliminates the $U(1)$
part of $A$ even if one starts with an $U(N)$ gauge
potential.

\smallskip

\noindent One also checks that the coupling of the Yang
Mills field $A$ with the fermions is equal to
$$
\lgl \psi ,D \psi \rgl \qquad \psi \in \Hc \, . \eqno
(2.3)
$$
The operator $D=D_0 +A+JAJ^*$ is given by
$$
D = e_a^{\mu} \, \g^a \left((\part_{\mu} + \om_{\mu}) \ot
1_N + 1 \ot \left( -{i\over 2} \, g_0 \, A_{\mu}^i \, T^i
\right)\right) \eqno (2.4)
$$
where $\om_{\mu}$ is the spin-connection on $M$:
$$
\om_{\mu} = {1\over 4} \, \om_{\mu}^{\ ab } \, \g_{ab}
$$
and $T^i $ are matrices in the adjoint representation
of SU(N) satisfying $\rm{Tr} (T^iT^j)=2\delta^{ij}$.
($\om_{\mu}^{ab}$ is related to the $e_{\mu}^a$ by
the vanishing of the covariant
derivative\footnote{$^{(*)}$}{\sevenrm We have limited
our considerations to torsion free spaces. The more
general case of torsion will be treated somewhere else.},
$$
\nb_{\mu} \, e_{\nu}^a = \part_{\mu} \, e_{\nu}^a -
\om_{\mu}^{ab} \, e_{\nu}^b - \G_{\mu \nu}^{\rho} \,
e_{\rho}^a = 0 \, . \eqno (2.5) 
$$
As the Christoffel connection
$$
\G_{\mu \nu}^{\rho} = {1\over 2} \, g^{\rho \s} (g_{\mu \s
, \nu} + g_{\nu \s , \mu} - g_{\mu \nu ,\s}) \eqno (2.6)
$$
is a given function of $g_{\mu \nu} = e_{\mu}^a
e_{\nu}^a$, equation (2.5) could be solved to express
$\om_{\mu}^{ab}$ as a function of $e_{\mu}^a$.)

\smallskip

\noindent It is a simple exercise to compute the square
of the Dirac operator given by (2.4) [10-11]. This can be
cast into the elliptic operator form [12]:
$$
P=D^2 =-(g^{\mu \nu} \, \part_{\mu} \, \part_{\nu} \cdot
\un + \Ab^{\mu} \, \part_{\mu} + \Bb) \eqno (2.7)
$$
where $\un$, $\Ab^{\mu}$ and $\Bb$ are matrices of the
same dimensions as $D$, and are given by:
$$
\displaylinesno{
\Ab^{\mu} = (2\om^{\mu} -\G^{\mu}) \ot 1_N - ig_0 \, 1_4 \ot
A^{\mu i} T^i &(2.8) \cr
\Bb = (\part^{\mu} \, \om_{\mu} + \om^{\mu} \, \om_{\mu} -
\G^{\nu} \om_{\nu} +R) \ot 1_N -ig_0 \, \om_{\mu} \ot
A^{\mu i} \, T^i \, . \cr
}
$$
In deriving (2.8) we have used equation (2.5) as well as
the following definitions and identities
$$
\displaylinesno{
[\part_{\mu} + \om_{\mu} , \part_{\nu} + \om_{\nu}] \eqv
{1\over 4} R_{\mu \nu}^{\ \ ab} (\om (e)) \g_{ab} \cr
e_{\rho}^a \, e_{\s}^b \, R_{\mu \nu}^{\ \ ab} (\om (e)) =
R_{\mu \nu \rho \s} (g) &(2.9) \cr
R_{\nu \rho \s}^{\ \ \ \mu} = \part_{\rho} \, \G_{\nu \s}^{\mu} -
\part_{\s} \, \G_{\nu \rho}^{\mu} + \G_{\rho \k}^{\mu} \,
\G_{\nu \s}^\k - \G_{\s \k}^{\mu} \, \G_{\nu \rho}^\k \cr
\G^{\mu} = g^{\nu \s} \, \G_{\nu \s}^{\mu} \cr
}
$$
we have also used the symmetries of the Riemann tensor to
prove that
$$
\g^{\mu \nu} R_{\mu \nu}^{\ \ ab} \, \g_{ab} = -2R \, . \eqno
(2.10)
$$
We shall now compute the spectral action
for this theory given by
$$
\Tr \chi \left( {D^2 \over m_0^2} \right) + (\psi ,D \, 
\psi) \eqno (2.11)
$$
where the trace $\Tr$ is the usual trace of operators in the
Hilbert space $\Hc$, and $m_0$ is a (mass) scale to be
specified. The function $\chi $ is chosen to be positive and
this has important consequences for the positivity of the
gravity action.

\smallskip

\noindent Using identities [12]:
$$
\Tr (P^{-s}) = {1\over \G (s)} \int_0^{\ify} t^{s-1} \Tr
e^{-tP} dt \qquad \Re (s) \geq 0 \eqno (2.12)
$$
and the heat kernel expansion for
$$
\Tr e^{-tP} \sm \sum_{n\geq 0} t^{{n-m \over d}} \int_M
a_n (x,P) \, dv (x) \eqno (2.13)
$$
where $m$ is the dimension of the manifold in $C^{\ify}
(M)$, $d$ is the order of $P$ (in our case $m=4$,
$d=2$) and $dv (x) = \sqrt g \, d^m \, x$ where $g^{\mu
\nu}$ is the metric on $M$ appearing in equation (2.7).

\smallskip

\noindent If $s=0,-1,\ldots$ is a non-positive integer
then $\Tr (P^{-s})$ is regular at this value of $s$ and is
given by
$$
\Tr (P^{-s}) = \Res \, \G (s) \mid_{s = {m-n \over d}} a_n
\, .
$$
From this we deduce that
$$
\Tr \chi (P) \sm \sum_{n\geq 0} f_n \, a_n (P) \eqno (2.14)
$$
where the coefficients $f_n$ are given by
$$
\displaylinesno{
f_0 = \int_0^{\ify} \chi (u) \, udu \ , \ f_2 =
\int_0^{\ify} \chi (u) \, du \ , \cr
f_{2(n+2)} = (-1)^n \, \chi^{(n)} (0) \ , \ n\geq 0 &(2.15)
\cr }
$$
and $a_n (P) = \int a_n (x,P) \, dv (x)$.

\medskip

The Seeley-de Witt coefficients $a_n (P)$ vanish for odd
values of $n$. The first three $a_n$'s for $n$ even are
[12]:
$$
\eqalignno{
a_0 (x,P) = & \, (4\pi)^{-m/2} \Tr (\un) \cr
a_2 (x,P) = & \, (4\pi)^{-m/2} \Tr \left(-{R \over 6}
\, \un +\Eb \right) \cr
a_4 (x,P) = & \, (4\pi)^{-m/2} {1 \over 360} \Tr ((-12
R;_{\mu} {}^{\mu} + 5R^2 - 2R_{\mu \nu} \, R^{\mu \nu}
\cr + & \, 2R_{\mu \nu \rho \s} \, R^{\mu \nu \rho \s})
\un - 60 R\Eb + 180 \Eb^2 + 60 \Eb;_{\mu} {}^{\mu} \cr
+ & \, 30 \Om_{\mu \nu} \, \Om^{\mu \nu})) & (2.16) \cr
}
$$
where $\Eb$ and $\Om_{\mu \nu}$ are defined by
$$
\eqalignno{
\Eb = & \, \Bb - g^{\mu \nu} (\part_{\mu} \, \om'_{\nu} +
\om'_{\mu} \, \om'_{\nu} - \G_{\mu \nu}^{\rho}
\, \om'_{\b}) \cr 
\Om_{\mu \nu} = & \, \part_{\mu} \, \om'_{\nu}
- \part_{\nu} \, \om'_{\mu} + [\om'_{\mu} \, \om'_{\nu}]
&(2.17) \cr 
\om'_{\mu} = & \, {1\over 2} \, g_{\mu \nu} (\Ab^{\nu} -
\G^{\nu} \cdot \un ) \, . \cr
}
$$
The Ricci and scalar curvature are defined by
$$
\eqalignno{
R_{\mu \rho} = & \, R_{\mu \nu}^{\ \ ab} \, e_b^{\nu}
\, e_{a\rho} \cr 
R = & \, R_{\mu \nu}^{\ \ ab} \, e_a^{\mu} \, e_b^{\nu}
\, . &(2.18) \cr
}
$$
We now have all the necessary tools to evaluate
explicitely the spectral action (2.11). Using equations
(2.8) and (2.16) we find :
$$
\eqalignno{
\Eb = & \, {1\over 4} R \ot \un_4 \ot \un_N + {i \over 4}
\, \g^{\mu \nu} \ot g F_{\mu \nu}^i \, T^i \cr
\Om_{\mu \nu} = & \, {1\over 4} R_{\mu \nu}^{ab} \,
\g_{ab} \ot 1_N - {i \over 2} \un_4 \ot g F_{\mu \nu}^i
\, T^i \, . &(2.19) \cr
}
$$
From the knowledge that the invariants of the heat
equation are polynomial functions of $R$, $R_{\mu \nu}$,
$R_{\mu \nu \rho \s}$, $\Eb$ and $\Om_{\mu \nu}$ and
their covariant derivatives, it is then evident from
equation (2.19) that the spectral action would not only
be diffeomorphism invariant but also gauge invariant. The
first three invariants are
then\footnote{$^{(*)}$}{\sevenrm Note that according to
our notations the scalar curvature $R$ is negative for
spheres.} 
$$ 
\eqalignno{
a_0 (P) = & \, {N \over 4\pi^2} \int_M \sqrt g \, d^4 x \cr
a_2 (P) = & \, {N\over 48\pi^2} \int_M \sqrt g  \, R
\, d^4 x &(2.20) \cr
a_4 (P) = & \, {1\over 16 \pi^2} \cdot {N \over 360} \int_M
d^4 x \, \sqrt g \biggl[(12 R;_{\mu} {}^{\mu} + 5R^2 -8R_{\mu
\nu} \, R^{\mu \nu} \cr
- & \, 7 R_{\mu \nu \rho \s} \, R^{\mu \nu \rho \s}) + {120
\over N} \, g^2 F_{\mu \nu}^i \, F^{\mu \nu i} \biggl] \, .
\cr
}
$$
For the special case where the dimension of the manifold
$M$ is four, we have a relation between the Gauss-Bonnet
topological invariant and the three possible curvature
square terms:
$$
R^* R^* = R_{\mu \nu \rho \s} \, R^{\mu \nu \rho \s} - 4
R_{\mu \nu} \, R^{\mu \nu} + R^2 \eqno (2.21)
$$
where $R^* R^* \eqv {1\over 4} \, \ve^{\mu \nu \rho \s} 
\, \ve_{\a \b \g \d} \, R_{\mu \nu}^{\a \b} \, R_{\rho
\s}^{\g \d}$. Moreover, we can change the expression for
$a_4 (P)$ in terms of $C_{\mu \nu \rho \s}$ instead of
$R_{\mu \nu \rho \s}$ where
$$
C_{\mu \nu \rho \s} = R_{\mu \nu \rho \s} - (g_{\mu
[\rho} R_{\nu \vert \s ]} - g_{\nu [\rho} R_{\mu \vert \s
]}) + {1\over 6} (g_{\mu \rho} \, g_{\nu \s} - g_{\mu \s}
\, g_{\nu \rho}) R \eqno (2.22)
$$
is the Weyl tensor. Using the identity:
$$
R_{\mu \nu \rho \s} \, R^{\mu \nu \rho \s} = C_{\mu \nu
\rho \s} \, C^{\mu \nu \rho \s} + 2R_{\mu \nu} \, R^{\mu
\nu} - {1\over 3} R^2 \eqno (2.23)
$$
we can recast $a_4 (P)$ into the alternative form:
$$
\eqalignno{
a_4 (p) = {N\over 48 \pi^2} \int d^4 x \, \sqrt g \biggl[
- & {3\over 20} \, C_{\mu \nu \rho \s} \, C^{\mu \nu \rho
\s} + {1\over 120} (11 R^* R^* + 12 R;_{\mu} {}^{\mu}) \cr
+ & \, {g^2 \over N} \, F_{\mu \nu}^i \, F^{\mu \nu i}
\biggl] &(2.24) \cr
}
$$
and this is explicitely conformal invariant. The Euler
characteristic $\chi_E$ (not to be confused with
the function $\chi $) is related to $R^* R^*$ by the
relation
$$
\chi_E = {1\over 32\pi^2} \int d^4 x \, \sqrt g \, R^* R^* \,
. \eqno (2.25)
$$
If we choose the function $\chi $ to be a cutoff function,
 i.e. $\chi (x)
=1$ for $x$ near $0$, then $\chi^{(n)} (0)$ is zero $\fl \,
n>0$ and this removes the non-renormalizable interactions. It
is, also possible to introduce a mass scale $m_0$ and consider
$\chi $ to be a function of the dimensionless variable $\chi
\left( {P \over m_0^2}\right)$. In this case terms coming
from $a_n (P)$, $n>4$ will be supressed by powers of ${1\over
m_0^2}$: 
$$
\eqalignno{
I_b =& \, {N\over 48\pi^2} \biggl[ 12 m_0^4 f_0 \int d^4
x \, \sqrt g + m_0^2 f_2 \int d^4 x \, \sqrt g \, R \cr
+ & \, f_4 \int d^4 x \, \sqrt g \biggl[ -{3\over 20}
\, C_{\mu \nu \rho \s} \, C^{\mu \nu \rho \s} + {1\over 10}
\, R;_{\mu} {}^{\mu} + {11 \over 20} \, R^* R^* \cr
+ & \, {g^2 \over N} \, F_{\mu \nu}^i \, F^{\mu \nu i}
\biggl] + 0 \left({1\over m_0^2}\right) \biggl] \, .
&(2.26) \cr 
}
$$
We shall adopt Wilson's view point of the renormalization
group approach to field theory [13] where the spectral action
is taken to give the {\it bare} action with bare quantities
$m_0$ and $g_0$ and with a cutoff scale $\Lb$ where the
theory is assumed to take a geometrical form. Introducing
the cutoff scale $\Lb$ will regularize the theory. The
perturbative expansion is then reexpressed in terms of {\it
renormalized} physical quanties. The fields also receive
wave function renormalization. Normalizing the Einstein and
Yang-Mills terms in the bare action we then have:
$$
\eqalignno{
{N \, m_0^2 \, f^2 \over 24\pi^2} = & \ {1\over \k_0^2} \eqv
{1\over 8\pi G_0} \cr
{f_4 \, g_0^2 \over 12\pi^2} = & \ 1 &(2.27) \cr
}
$$
and (2.26) becomes:
$$
\eqalignno{
I_b = &\int d^4 x \, \sqrt g 
\biggl[ {1\over 2\k_0^2} \, R +e_0 \cr
 + & \,  a_0 \, C_{\mu \nu \rho \s}
\, C^{\mu \nu \rho \s} + c_0 \, R^* \, R^* + d_0 \, R;_{\mu}
{}^{\mu}  + {1\over 4} \, F_{\mu \nu}^i \, F^{\mu \nu i}
\biggl] &(2.28)\cr
}
$$
where
$$
\eqalignno{
a_0 = & \ {-3N \over 80} \ {1\over g_0^2} &(2.29) \cr
c_0 = & \ -{2\over 3} \, a_0 \cr
d_0 = & \ -{11 \over 3} \, a_0 \cr
e_0 = & \ {N \, m_0^4 \over 4\pi^2} \, f_0 \, . \cr
}
$$
The renormalized action receives counterterms of the same
form as the bare action but with physical parameters
$k,a,c,d$, and the addition of one new term [14]
$$
\int d^4 x \, \sqrt g \ (b \, R^2) \, . \eqno (2.30)
$$
This adds one further boundary condition for the equations
(2.29):
$$
b_0 = 0 \, .
$$
The renormalized fermionic action $(\psi ,D\psi)$ keeps the
same form as the bare fermionic action.

\smallskip

\noindent The renormalization group equations will yield
relations between the bare quantities and the physical
quantities with the addition of the cutoff scale $\Lb$.
Conditions on the bare quantities would translate into
conditions on the physical quantities. In the present
example only the gauge coupling $g(\Lb)$ and Newton's
constant will have measurable effects. The dependence of
$\k_0$ on $\k$ and the other physical quantities is such that
$\k_0^{-2} - \k^{-2}$ contains terms proportional to the
cutoff scale. As $\k^2$ must be identified with $8\pi G$ at
low energy it is clear that both $\k_0^{-1}$ and $\Lb$ could
be as high as the Planck scale\footnote{$^{(*)}$}{\sevenrm We like to
thank A. Tseytlin for correspondence on this point.}

\smallskip

\noindent The renormalization group equations of this system
were studied by Fradkin and Tseytlin [15] and is known to be
renormalizable, but non-unitary [14] due to the presence of
spin-two ghost (tachyon) pole near the Planck mass. We shall
not worry about non-unitarity (see, however, reference 16),
because in our view at the Planck energy the manifold
structure of space-time will break down and one must have a
completely finite theory where only the part of the Hilbert
space given by $\chi (D^2) \Hc$ enters. The algebra
$\Ac$ becomes finite dimensional in such a way that all
symmetries of the continuum (in some approximation) would be
admitted.

\smallskip

\noindent In the limit of flat space-time we have $g_{\mu
\nu} = \d_{\mu \nu}$ and the action (2.11) becomes
(adopting the normalizations (2.29)):
$$
{1\over 4} \, F_{\mu \nu}^i \, F^{\mu \nu i} + (\psi , D
\, \psi) \eqno (2.31)
$$
where we have droped the constant term. This action is
known to have $N =1$ global supersymmetry. In reality we
can also obtain the $N=2$ and $N=4$ super Yang-Mills
actions by taking the appropriate Dirac operators in six
and ten dimensions respectively [17].

\vglue 1cm

\noindent {\bf 3. Spectral action for the standard model.}

\medskip

Having illustrated the computation of our spectral action
for the Einstein-Yang-Mills system we now
address the realistic case of obtaining action (1.2) for
the Einstein-Standard model system. 

\smallskip

\noindent We first briefly summarize the spectral triple
$(\Ac ,\Hc ,D)$ associated with the spectrum of the standard
model. A complete treatment can be found in [4,6].

\smallskip

\noindent The geometry is that of a 4-dimensional smooth
Riemannian manifold with a fixed spin structure times a
discrete geometry. The product geometry is given by the rules
$$
\Ac = \Ac_1 \ot \Ac_2 \ , \ \Hc = \Hc_1 \ot \Hc_2 \ , \ D =
D_1 \ot 1 + \g_5 \ot D_2 \eqno (3.1)
$$
where $\Ac_1 = C^{\ify} (M)$, $D_1 = {\part \!\!\! /}_M$ the
Dirac operator on $M$, $\Hc_1 = L^2 (M,S)$ and the discrete
geometry $(\Ac_2 ,\Hc_2 ,D_2)$ will now be described. The
algebra $\Ac_2$ is the direct sum of the real involutive
algebras $\Cb$ of complex numbers, $\Hb$ of quaternions, and
$M_3 (\Cb)$ of $3\ts 3$ matrices. $\Hc_2$ is the Hilbert space
with basis the elementary fermions, namely the quarks $Q$,
leptons $L$ and their charge conjugates, where
$$
Q = \left( \matrix{
u_L \cr d_L \cr d_R \cr u_R \cr
} \right) \ , \ L = \left( \matrix{
\nu_L \cr e_L \cr e_R \cr
} \right) \eqno (3.2)
$$
and we have omitted family indices for $Q$ and $L$ and
colour index for $Q$. The antilinear isometry $J = J_2$ in
$\Hc_2$ exchanges a fermions with its conjugate.

\smallskip

\noindent The action of an element $a = (\lb ,q, m) \,
\in \Ac_2$ in $\Hc_2$ is given by:
$$
a \, Q = \left( \matrix{
q \, \left( \matrix{ u_L \cr d_L \cr } \right) \cr
\bar{\lb} \, d_R \cr
\lb \, u_R \cr
} \right) \eqno (3.3)
$$
where $q= \left( \matrix{ \a &\b \cr -\bar{\b} &\bar{\a} \cr
} \right)$ is a quaternion. A similar formula holds for
leptons. The action on conjugate particles is:
$$
\eqalignno{
a \, \bar L = & \ \lb \, \bar L \cr
a \, \bar Q = & \ m \, \bar Q \, . &(3.4) \cr
}
$$
For the operator $D_2$ we take $D_2 = \left( \matrix{ Y &0
\cr 0 &\bar Y \cr } \right)$ where $Y$ is a Yukawa coupling
matrix of the form
$$
Y = Y_q \ot 1_3 \op Y_{\ell} \eqno (3.5)
$$
with
$$
Y_q = \left( \matrix{
0_2 &k_0^d \ot H_0 &k_0^u \ot \wt{H}_0 \cr
(k_0^d)^* \ot H_0^* &(k_0^u)^* \ot \wt{H}_0^* &0_2 \cr
} \right) \eqno (3.6)
$$
$$
Y_{\ell} = \left( \matrix{
0_2 &k_0^e \ot H_0 \cr
k_0^{e*} \ot H_0^* &0 \cr
} \right) \, .
$$
The matrices $k^d$, $k^u$ and $k^e$ are $3\ts 3$ family
mixing matrices and 
$$
H_0 = \mu \left( \matrix{ 0 \cr 1 \cr}\right) \ , \ \wt{H}_0
= i\s_2 \, H_0 \, . 
$$
The parameter $\mu$ has the dimension of mass.

\smallskip

\noindent The choice of the Dirac operator and the action of
$\Ac_2$ in $\Hc_2$ comes from the restrictions that these must
satisfy:
$$
\displaylinesno{
J^2 = 1 \ , \ [J ,D_2] = 0 \ , \ [a,Jb^* J^{-1}] = 0 \cr
[[D,a] , J b^* J^{-1}]=0 \quad \fl \, a,b \, . &(3.7)
}
$$
The next step is to compute the inner fluctuations of the
metric and thus the operators of the form: $A = \Si \, a_i
[D,b_i]$. This with the self-adjointness condition $A=A^*$
gives a $U(1)$, $SU(2)$ and $U(3)$ gauge fields as well as a
Higgs field. The computation of $A+ JAJ^{-1}$ removes a
$U(1)$ part from the above gauge fields (such that the full
matrix is traceless) (for derivation see [4]). The Dirac
operator $D_q$ that takes the inner fluctuations into account
is given by the $36 \ts 36$ matrix (acting on the 36 quarks)
(tensored with Clifford algebras)
$$
\eqalignno{
D_q & = \cr
&\left[ \matrix{
\g^{\mu} \ot \left(D_{\mu} \ot 1_2 -{i\over 2} g_{02}
A_{\mu}^{\a} \s^{\a} -{i \over 6} g_{01} B_{\mu} \ot 1_2
\right) \ot 1_3 , \ \g_5 \ot k_0^d \ot H , \ \g_5 \ot k_0^u
\ot \wt H \cr
\g_5 \ot k_0^{d*} \ot H^* , \qquad \qquad \g^{\mu} \ot
\left( D_{\mu} + {i \over 3} g_{01} B_{\mu} \right) \ot 1_3 ,
\qquad \qquad \qquad \qquad 0 \hfill \cr 
\g_5 k_0^{u*} \wt{H}^* , \qquad \qquad \qquad \qquad \qquad
0 , \qquad \qquad \g^{\mu} \ot \left( D_{\mu} -{2i \over 3}
g_{01} B_{\mu} \right) \ot 1_3 \hfill \cr } \right] \ot 1_3
\cr 
& + \g^{\mu} \ot 1_4 \ot 1_3 \ot \left( -{i \over 2} \,
g_{03} \, V_{\mu}^i \, \lb^i \right) &(3.8) \cr
}
$$
where $\s^{\a}$ are Pauli matrices and $\lb^i$ are
Gell-mann matrices satisfying
$$
\Tr (\lb^i \lb^j) = 2\d^{ij} \, . \eqno (3.9)
$$
The vector fields $B_{\mu}$, $A_{\mu}^{\a}$ and
$V_{\mu}^i$ are the $U(1)$, $SU(2)_w$ and $SU(3)_c$ gauge
fields with gauge couplings $g_{01}$, $g_{02}$ and $g_{03}$.
The differential operator $D_{\mu}$ is given by
$$
D_{\mu} = \part_{\mu} + \om_{\mu} \eqno (3.10)
$$
and $\g^{\mu} = e_a^{\mu} \, \g^a$. The scalar field $H$ is
the Higgs doublet, and $\wt H$ is the $SU(2)$ conjugate
of $H$:
$$
\wt H = (i \s^2 H) \, . \eqno (3.11)
$$
We note that although $H_0$ was introduced in the definition
of $D_2$ it is absorbed in the field $H$.

\smallskip

\noindent It is a simple exercise to see that the action
for the fermionic quark sector is given by
$$
(Q,D_q \, Q) \, . \eqno (3.12)
$$
The Dirac operator acting on the leptons, taking inner
fluctuations into account is given by the  $9\ts 9$ matrix
(tensored with Clifford algebra matrices): 
$$
D_{\ell} = \left[ \matrix{
\g^{\mu} \ot \left( D_{\mu} -{i \over 2} g_{02} A_{\mu}^{\a}
\s^{\a} + {i \over 2} g_{01} B_{\mu} \ot 1_2 \right) \ot 1_3
\qquad \qquad \g_5 \ot k_0^e \ot H \cr
\g_5 \ot k_0^{*e} \ot H^* \hfill \qquad \qquad \qquad
\qquad \qquad \g^{\mu} \ot (D_{\mu} + ig_{01} B_{\mu}) \ot
1_3 \cr 
} \right] \, . \eqno (3.13)
$$
Again the leptonic action have the simple form
$$
(L,D_{\ell} \, L) \, . \eqno (3.14)
$$
According to our universal formula (1.28) the spectral
action for the standard model is given by:
$$
\Tr [ \chi (D^2 / m_0^2 )] + (\psi ,D \psi) \eqno
(3.15) 
$$
where $(\psi ,D\psi)$ will include the quark sector (3.12)
and the leptonic sector (3.14). Calculating the bosonic part
of the above action follows the same lines as in the previous
section. The steps that lead to the result are given in the
appendix.

\medskip

The bosonic action is
$$
\eqalignno{
I = & \, {9m_0^4 \over \pi^2} {5 \over 4} \,
f_0 \int d^4 x \, \sqrt g \cr
+ & \, {3m_0^2 \over 4\pi^2} f_2 \int d^4 x \, \sqrt g
\left[ {5\over 4} \, R-2 y^2 H^* H \right] \cr
+ & \, {f_4 \over 4\pi^2} \int d^4 x \, \sqrt g \biggl[
{1\over 40} \, {5\over 4} \, (12 R;_{\mu} {}^{\mu} +
11R^* R^* - 18 C_{\mu \nu \rho \s} \, C^{\mu \nu \rho \s})
\cr
+ & \, 3 y^2 \left( D_{\mu} \, H^* D^{\mu} H -{1\over 6} R
\, H^* H \right) \cr
+ & \, g_{03}^2 \, G_{\mu \nu}^i \, G^{\mu \nu i} +  g_{02}^2 \,
F_{\mu \nu}^{\a} \, F^{\mu \nu \a} \cr
+ & \, {5\over 3} \, g_{01}^2
\,  B_{\mu \nu} \, B^{\mu \nu} \cr
+ & \, 3 z^2 (H^* H)^2 - y^2 (H^* H);_{\mu} {}^{\mu} \biggl]
+ 0 \left( {1\over m_0^2}\right) &(3.16) \cr
}
$$
where we have denoted
$$
\eqalignno{
y^2 = & \Tr \left( \vert k_0^d \vert^2 + \vert k_0^u \vert^2
+ {1\over 3} \vert k_0^e \vert^2 \right) \cr
z^2 = & \Tr \left( (\vert k_0^d \vert^2 + \vert k_0^u
\vert^2)^2 + {1\over 3} \vert k_0^e \vert^4 \right) &(3.17)
\cr
D_{\mu} \, H = & \, \part_{\mu} \, H - {i \over 2} \, g_{02}
\, A_{\mu}^{\a} \, \s^{\a} H - {i \over 2} \, g_{01}
\, B_{\mu} \, H \, . \cr
}
$$
Normalizing the Einstein and Yang-Mills terms gives:
$$
\eqalignno{
{15m_0^2 \, f_2 \over 4\pi^2} = & \ {1\over \kappa_0^2} \cr
{g_{03}^2 \, f_4 \over \pi^2} = & \ 1 &(3.18) \cr
g_{03}^2 = & \ g_{02}^2 = {5\over 3} \, g_{01}^2 \, . \cr
}
$$
Relations (3.18) among the gauge coupling constants coincide
with those coming from $SU(5)$ unification.

\smallskip

\noindent To normalize the Higgs fields kinetic energy we
have to rescale $H$ by:
$$
H \ra {2\over 3} \, {g_{03} \over y} \, H \, . \eqno (3.19)
$$
This transforms the bosonic action (3.16) to the form:
$$
\eqalignno{
I_b = & \ \int d^4 x \, \sqrt g \ \biggl[ {1\over 2\kappa_0^2} \,
R - \mu_0^2 (H^* H) + a_0 \, C_{\mu \nu \rho \s} \, C^{\mu
\nu \rho \s} \cr
+ & \ b_0 \, R^2 + c_0 \, {}^* R {}^* R + d_0 \, 
R;_{\mu} {}^{\mu} \cr
+ & \ e_0 + {1\over 4} \, G_{\mu \nu}^i \, G^{\mu \nu i} +
{1\over 4} \, F_{\mu \nu}^{\a} \, F^{\mu \nu \a} \cr
+ & \ {1\over 4} \, B_{\mu \nu} \, B^{\mu \nu} + \vert
D_{\mu} \, H\vert^2 - \xi_0 \, R \vert H \vert^2 + \lb_0
(H^* H)^2 \biggl] &(3.20) \cr
}
$$
where
$$
\eqalignno{
\mu_0^2 = & \ {4\over 3\kappa_0^2} \cr
a_0 = & \ -{9 \over 8g_{03}^2} \cr
b_0 = & \ 0 &(3.21) \cr
c_0 = & \ -{11 \over 18} \, a_0 \cr
d_0 = & \ -{2\over 3} \, a_0 \cr
e_0 = & \ {45 \over 4\pi^2} \, f_0 \, m_0^4 \cr
\lb_0 = & \ {4\over 3} \, g_{03}^2 \, {z^2 \over y^4} \cr
\xi_0 = & \ {1\over 6} \, . \cr
}
$$
As explained in the last section this action has to be taken
as the {\it bare action} at some cutoff scale $\Lb$. The
renormalized action will have the same form as (3.20) but
with the bare quantities $\k_0$, $\mu_0$, $\lb_0$, $a_0$ to
$e_0$ and $g_{01}$, $g_{02}$, $g_{03}$ replaced with
physical quantities.

\smallskip

\noindent Relations between the bare gauge coupling
constants as well as equations (3.19) have to be imposed as
boundary conditions on the renormalization group equations
[13]. The bare mass of the Higgs field is related to the
bare value of Newton's constant, and both have quadratic
divergences in the limit of infinite cutoff $\Lb$. The
relation between $m_0^2$ and the physical quantities is:
$$
m_0^2 = m^2 \left( 1 + {\left( {\Lb^2 \over m^2} -1 \right)
\over 32\pi^2} \ \left( {9\over 4} \, g_2^2 + {3\over 4} \,
g_1^2 + 6\lb - 6k_t^2 \right)\right) + 0 \left( \ln \, {\Lb^2
\over m^2} \right) + \ldots \eqno (3.22)
$$
For $m^2 (\Lb)$ to be small at low-energies $m_0^2$ should
be tuned to be proportional to the cutoff scale according to
equation (3.22).

\smallskip

\noindent Similarly the bare cosmological constant is
related to the physical one (which must be tuned to zero at
low energies):
$$
e_0 = e + {\Lb^4 \over 32\pi^2} \, (62) + \ldots \eqno (3.23)
$$
where 62 is the difference between the fermionic degrees of
freedom (90) and the bosonic ones (28).

\smallskip

\noindent There is also a relation between the bare scale
$\k_0^{-2}$ and the physical one $\k^{-2}$ which is similar to
equation (3.20) (but with all one-loop contributions coming
with the same sign) which shows that $\k_0^{-1} \sim m_0$ and
$\Lb$ are of the same order as the Planck mass.

\smallskip

\noindent The renormalization group equations for the gauge
coupling constants are: 
$$
\eqalignno{
{dg_1 \over dt} = & \, {1\over 16\pi^2} \left( {41 \over
6} \right) g_1^3 \cr
{dg_2 \over dt} = & \, {1\over 16\pi^2} \left( -{19 \over
6} \right) g_2^3 &(3.24) \cr
{dg_3 \over dt} = & \, {1\over 16\pi^2} (-7) g_3^3 \cr
}
$$
where $t=\ln \, \mu$, $\mu$ being the running scale.

\medskip

Solutions to equations (3.22) are known from the $SU(5)$
case and are given by [19]
$$
\eqalignno{
\a_1^{-1} (M_Z) = & \, \a_1^{-1} (\Lb) + {41 \over 12\pi}
\, \ln \, {\Lb \over M_Z} \cr
\a_2^{-1} (M_Z) = & \, \a_2^{-1} (\Lb) -{19 \over 12\pi}
\, \ln \, {\Lb \over M_Z} &(3.25) \cr
\a_3^{-1} (M_Z) = & \, \a_3^{-1} (\Lb) - {42 \over 12\pi}
\, \ln  \, {\Lb \over M_Z} \cr
}
$$
where $\a_i = {g_i^2 \over 4\pi}$, $i=1,2,3$ and $M_z$ is
the mass of the $Z$ vectors. At the scale $\Lb$ we have to
impose the boundary conditions (3.18): 
$$
\a_3 (\Lb) = \a_2 (\Lb) = {5\over 3} \, \a_1 (\Lb) \, .
\eqno (3.26)
$$
Using equations (3.27) and (3.28) one easily find:
$$
\eqalignno{
\sin^2 \t_w = & \, {3\over 8} \left[ 1-{109 \over 18\pi}
\, \a_{em} \, \ln \, {\Lb \over M_Z} \right] \cr
\ln \, {\Lb \over M_Z} = & \, {2\pi \over 67}
(3\a_{em}^{-1} (M_Z) - 8\a_3^{-1} (M_Z)) \, . &(3.27) \cr
}
$$
The present experimental values for $\a_{em}^{-1} (M_Z)$
and $\a_3 (M_Z)$ are
$$
\displaylinesno{
\a_{em}^{-1} (M_Z) = 128.09 \cr
0.110 \leq \a_3 (M_Z) \leq 0.123 \, . &(3.28)
}
$$
These values lead to
$$
\displaylinesno{
9.14 \ts 10^{14} \leq \Lb \leq 4.44 \ts 10^{14}
(\Gev) \cr
0.206 \leq \sin^2 \t_w \leq 0.210 \, . &(3.29) \cr
}
$$
Therefore the bare action we obtained and associated with
the spectrum of the standard model is consistent with
experimental data provided the cutoff scale is taken to be
$\Lb \sim 10^{15} \Gev$. There is, however, a slight
disagreement (10\%) between the predicted value of $\sin^2
\, \t_w$ and the experimental value of 0.2325 known to a very
high precision. It is a remarkable fact that starting from
the spectrum of the standard model at low-energies, and
assuming that this spectrum does not change, one can get the
geometrical spectral action which holds at very
high-energies and consistent within ten percent with
experimental data. This can be taken that at higher energies
the noncommutative nature of space-time reveals itself and
shows that the effective theory at the scale $\Lb$ have a
higher symmetry. The other disagreement is that the gravity
sector requires the cutoff scale to be of the same order as
the Planck scale while the condition on gauge coupling
constants give $\Lb \sim 10^{15} \Gev$. The gravitational coupling
$G$ runs with $\Lb$ due to the matter interactions. This dictates
that it must be of the order $\Lb^{-2}$ and gives a large value
for Newton's constant.  These results must be taken as an
indication that the spectrum of the standard model has to be
altered as we climb up in energy. The change may happen at
low energies (just as in supersymmetry which also pushes the
cutoff scale to $10^{16} \Gev$) or at some intermediate
scale. Incidently the porblem that Newton's constant is 
coming out to be too large is also present  in string
theroy where  also a unification of gauge couplings and
Newton's constant occurs [20].
Ultimately one would hope that modification of the
spectrum will increase the cutoff scale nearer to the Planck
mass as dictated by gravity.

\medskip

There is one further relation in our theory between the
$\lb (H^* H)^2$ coupling and the gauge couplings to be
imposed at the scale $\Lb$ [21]:
$$
\lb_0 = {4\over 3} \, g_{03}^2 \, {z^2 \over y^4} \, . \eqno
(3.30) 
$$
This relation could be simplified if we assume that the
top quark Yukawa coupling is much larger than all the
other Yukawa couplings. In this case equation (3.30)
simplifies to
$$
\lb (\Lb) = {16\pi \over 3} \a_3 (\Lb) \, . \eqno (3.31)
$$
Therefore the value of $\lb$ at the
unification scale is $\lb_0 \sm 0.402$ showing that one
does not go outside the perturbation domain. In reality,
equation (3.31) could be used, together with the RG
equations for $\lb$ and $k_t$ to determine the Higgs mass
at the low-energy scale $M_Z$ [22]:
$$
\eqalignno{
{d\lb \over dt} = & \, 4\lb \g + {1\over 8\pi^2} (12\lb^2
+B) \cr
{dk_t \over dt} = & \, {1\over 32\pi^2} \left[ 9k_t^3 -
\left( 16g_3^2 + {9\over 2} \, g_2^2 + {17 \over 6}
\, g_1^2 \right) k_t \right] &(3.32) \cr
}
$$
where
$$
\eqalignno{
\g = & \, {1\over 64\pi^2} (12k_t^2 - 9 g_t^2 - 3g_1^2) \cr
B = & \, {3\over 84\pi^2} \left( {1\over 16} (3g_2^4 + 2
g_1^2 \, g_2^2 -g_1^4) -k_t^4 \right) \, . &(3.33) \cr
}
$$
These equations have to be integrated numerically [21].
One can get a rough estimate on the Higgs mass from the
triviality bound\footnote{$^{(*)}$}{\sevenrm We like to
thank M. Lindner for explanations on this point.} on the
$\lb$ couplings. For $\Lb \sm 10^{15} \Gev$ as given in
equation (3.29) the limits are
$$
160 < m_H < 200 \Gev \, . \eqno (3.34)
$$

This together with the boundary condition (3.31) gives a
mass of the Higgs near the lower bound of $160 \Gev$. The
exact answer can be only determined by numerical
integration, but this of course cannot be completely
trusted as the predicted value for $\sin^2 \t_w$ is off
by ten percent. It can, however, be taken as an
approximate answer and in this respect one can say that
the Higgs mass lies in the interval $160 - 180 \Gev$. We
expect this prediction to be correct to the same precision
as that of $\sin^2 \t_w$ in (3.29).

\smallskip

\noindent In reality we can perform the same analysis for
the gravitational sector to determine the dependence of
$\k_0$, $a_0$, $b_0$, $c_0$, $d_0$ and $e_0$ on the physical
quantities and the effect of the boundary conditions (3.19)
on them. This, however, will not have measurable
consequences and will not be pursued here.

\vglue 1cm

\noindent {\bf 4. Conclusions.}

\medskip

The basic symmetry for a noncommutative
space $(\Ac ,\Hc ,D)$ is $\Aut (\Ac)$. This symmetry
includes diffeomorphisms and internal symmetry
transformations. The bosonic action is a spectral
function of the Dirac operator while the
fermionic action takes the simple linear form $(\psi
,D\psi)$ where $\psi$ are spinors defined on the Hilbert
space. Applying this principle to the simple case where
the algebra is $C^{\ify} (M) \ot M_n (\Cb)$ with a Hilbert
space of fermions in the adjoint representation one finds
that the bosonic action contains the Yang-Mills, Einstein
and Weyl actions. This action is to be interpreted as the
bare Wilsonian action at some cutoff scale $\Lb$. The same
principle when applied to the less trivial noncommutative
geometry of the standard model gives the standard model
action coupled to Einstein and Weyl gravity plus higher order
non-renormalizable interactions supressed by powers of the
inverse of the mass scale in the theory. One also gets a mass
term for the Higgs field. This bare mass is of the same order
as the cutoff scale and this is related to the fact that there
are quadratic divergences associated with the Higgs mass in
the standard model. There are some relations between the
bare quantities. The renormalized action will have the same
form as the bare action but with physical quantities
replacing the bare ones (except for an $R^2$ term which is
absent in the bare action due to the scale invariance of
the  $a_4$ term associated with the square of the Dirac
operator). The relations among the bare quantities must be
taken as boundary conditions on the renormalization group
equations governing the scale dependence of the physical
quantities.

\medskip

In particular there are relations among the gauge
couplings coinciding with those of $SU(5)$ (or any gauge
group containing  $SU(5)$ and also between
the Higgs couplings to be imposed at some scale. These
relations give a unification scale (or cutoff scale) of order
$\sim 10^{15} \Gev$ and a value for $\sin^2 \t_w \sim 0.21$
which is off by ten percent from the true value. We also have
a prediction of the Higgs mass in the interval $160-180 
\Gev$. This can be taken as an indication that the
noncommutative structure of space-time reveals itself at
such high scale where the effective action has a geometrical
interpretation.

\medskip

The slight disagreement with experiment
indicates that the spectrum of the standard model could
not be extrapolated to very high energies without adding
new particles necessary to change the RG equations of the
gauge couplings. One possibility could be supersymmetry,
but there could be also less drastic solutions. It might
be tempting by changing the spectrum to push the unification
scale up nearer to the Planck scale a situation
which is also present in string theory.

\medskip

In summary, we have succeeded in finding a universal
action formula that unified the standard model with the
Einstein action. This necessarily involved an
extrapolation from the low-energy sector to $10^{15} \Gev$,
assuming no new physics arise. Our slight disagreement for the
prediction of $\sin^2 \t_w$ and for the low value of the
unification scale seems to imply that the spectrum of the
standard model must be modified either at low-energy or at an
intermediate scale. There is also the possibility that by
formulating the theory at very high energies, the concept of
space-time as a manifold breaks down and the
noncommutativity of the algebra must be extended to include
the manifold part. One expects that the algebra $A$ becomes
a finite dimensional algebra. Finally, we hope that our
universal action formula should be applicable to many
situations of which the most important could be
superconformal field theory. Work along these ideas is now in
progress.

\vglue 1cm

\noindent {\bf Acknowledgments.} A.H.C. would like to
thank J\"urg Fr\"ohlich for very useful discussions and
I.H.E.S. for hospitality where part of this work was done.

\vfill\eject

\noindent {\bf Appendix.}

\medskip

To derive a general formula for $\Tr \chi (D^2 /\Lb^2)$ we
must evaluate the heat kernel invariants $a_n (x,P)$ for a
Dirac operator of the form
$$
D = \left( \matrix{
\g^{\mu} (D_{\mu} \cdot \un_N + A_{\mu}) &\g_5 S \cr
\g_5 S &\g^{\mu} (D_{\mu} \ot \un_N + A_{\mu}) \cr
} \right) \, . \eqno (A.1)
$$
Evaluating $D^2$ we find that
$$
\Ab^{\mu} = ((2\om^{\mu} -\G^{\mu}) \ot 1_N + 2A^{\mu})
\ot 1_2 \eqno (A.2)
$$
$$
\eqalignno{
\Bb = & \, (\part^{\mu} \, \om_{\mu} + \om^{\mu}
\, \om_{\mu} - \G^{\mu} \, \om_{\mu} +R) \ot 1_N +
2A^{\mu} \cdot \om_{\mu} \cr 
+ & \, \left(\part^{\mu} + \om^{\mu}
-\G^{\mu}) \, A_{\mu} \, -{1 \over 2} \, \g^{\mu
\nu} \, F_{\mu \nu} + A^{\mu} \, A_{\mu} -S^2 \right) \ot
1_2 \cr 
- & \, \g^{\mu} \, \g_5 (D_{\mu} S + [A_{\mu},S]) \ot
\left( \matrix{ 0 &1 \cr 1 &0 \cr} \right) \, . &(A.3)
}
$$
From this we can construct $\Eb$ and $\Om_{\mu \nu}$:
$$
\eqalignno{
\Eb = & \left( {1\over 4} R \ot 1_N -{1\over 2} \, \g^{\mu
\nu} \, F_{\mu \nu} -S^2 \right) \ot 1_2 \cr
- & \, \g^{\mu} \, \g_5 (D_{\mu} S +[A_{\mu},S])
\ot \left( \matrix{ 0 &1 \cr 1 &0 \cr} \right) &(A.4) \cr
}
$$
$$
\Om_{\mu \nu} = \left( {1\over 4} \, R_{\mu \nu}^{\ \ ab} \,
\g_{ab} \ot 1_N + \g^{\mu \nu} F_{\mu \nu}\right) \ot 1_2
\, . \eqno (A.5)
$$
From this we deduce that
$$
\eqalignno{
a_0 (x,P) = & \, {\Lb^4 \over 4\pi^2} \int \sqrt g \, d^4
x \, \Tr (1) \cr
a_2 (x,P) = & \, {\Lb^2 \over 4\pi^2} \int \sqrt g \, d^4
x \, \left[ {R\over 12} \Tr (1) -2\Tr (S^2) \right] \cr
a_4 (x,P) = & \, {1\over 4\pi^2} \int \sqrt g \, d^4 x \,
\biggl[ {\Tr (1) \over 360} \left( 3R;_{\mu} {}^{\mu} -
{9\over 2} C_{\mu \nu \rho \s} \, C^{\mu \nu \rho \s} +
{11\over 4} R^* R^* \right) \cr
+ & \, \Tr \left((D_{\mu} S +[A_{\mu}, S])^2 - {R\over 6}
S^2 \right) \cr
- & \, {1\over 6} \Tr F_{\mu \nu} \, F^{\mu \nu} + \Tr S^4
-{1\over 3} \Tr (S^2);_{\mu} {}^{\mu} \biggl] \, . &(A.6)
\cr 
}
$$
Applying these formulas to the Dirac operator of the
quark sector we can obtain the same answer as from an
explicit calculation by replacing in the previous formulas:
$$
\eqalignno{
\Tr (1) \ra & \, 36 \cr
\Tr S^2 \ra & \, 3\Tr (\vert k_0^d \vert^2 + \vert k_0^u
\vert^2) H^* H \cr
\Tr S^4 \ra & \, 3\Tr \vert (\vert k_0^d \vert^2 + \vert
k_0^u \vert^2)^2 \vert (H^* H)^2 \cr
A_{\mu} \ra & \left( \matrix{
\left( -{i\over 2} g_{02} A_{\mu}^{\a} \s^{\a} -{i\over 6}
g_{01} B_{\mu} \cdot \un_2 \right) \cr
&{1\over 3} g_{01} B_{\mu} \cr
&&-{2i \over 3} g_{01} B_{\mu} \cr
}\right) \ot 1_3 \ot 1_3 \cr
+ & \, 1_4 \ot 1_3 \ot \left( -{i \over 2} \, g_{03}
\, V_{\mu}^i \, \lb^i \right) \, . &(A.7) \cr
}
$$
Then 
$$
-{1\over 6} \Tr F_{\mu \nu} \, F^{\mu \nu} = {3\over 4} \,
g_{02}^2 \, F_{\mu \nu}^{\a} \, F^{\mu \nu \a} + {11 \over 2}
\, g_{01}^2 \, B_{\mu \nu} \, B^{\mu \nu} + g_{03}^2 \, G_{\mu
\nu}^i \, G^{\mu \nu i} \, . \eqno (A.8)
$$
In the leptonic sector, we make the replacements:
$$
\eqalignno{
\Tr (1) \ra & \, 9 \cr
\Tr S^2 \ra & \Tr \vert k_0^e \vert^2 H^* H \cr
\Tr S^4 \ra & \Tr \vert k_0^e \vert^4 (H^* H)^2 \cr
-{1\over 6} \Tr F_{\mu \nu} \, F^{\mu \nu} \ra & \,
{1\over 3} \left( {3\over 4} \, g_{02}^2 \, F_{\mu \nu}^{\a}
\, F^{\mu \nu \a} + {11\over 2} \, g_{01}^2 \, B_{\mu \nu}
\, B^{\mu \nu} + g_{03}^2 \, G_{\mu \nu}^i \, G^{\mu \nu i}
\right) \, . &(A.9) \cr 
}
$$

\vfill\eject

\noindent {\bf References.}

\medskip

\item{[1]}  A. Connes, {\it Publ. Math. IHES} {\bf 62},
44 (1983); Noncommutative Geometry (Academic Press,
New York 1994).

\item{[2]} A. Connes and J. Lott, {\it Nucl. Phys. Proc.
Supp.} {\bf B18}, 295 (1990); proceedings of 1991
Carg\`ese Summer Conference, edited by J. Fr\"ohlich et
al (Plenum, New York 1992).

\item{[3]} D. Kastler, {\it Rev. Math. Phys.} {\bf 5}, 477
(1993).

\item{[4]} A. Connes, ``Gravity Coupled with Matter and the
Foundation of Noncommutative Geometry'', hep-th/9603053.

\item{[5]} A.H. Chamseddine, G. Felder and J. Fr\"ohlich,
{\it Comm. Math. Phys.} {\bf 155}, 109 (1993); A.H.
Chamseddine, J. Fr\"ohlich and O. Grandjean, {\it J.
Math. Phys.} {\bf 36}, 6255 (1995).

\item{[6]} A. Connes, {\it J. Math. Phys.} {\bf 36}, 6194
(1995).

\item{[7]} B. De Witt, {\it Dynamical Theory of Groups and
Fields} (New York, Gordon and Breach 1965).

\item{[8]} S. Adler in {\it The high energy limit}, Erice
lectures edited by A. Zichichi (Plenum, New York 1983).

\item{[9]} H. Grosse, C. Klimcik and P. Presnajder,
``On Finite 4D Quantum Field Theory in Noncommutative Geometry'',
hep-th/9602115.

\item{[10]} D. Kastler, {\it Comm. Math. Phys.} {\bf 166},
633 (1995).

\item{[11]} W. Kalau and M. Walze, {\it J. Geom. Phys.}
{\bf 16}, 327 (1995).

\item{[12]} P. Gilkey, Invariance theory, the heat
equation and the Atiyah-Singer index theorem, (Publish or
Perish, Dilmington, 1984).

\item{[13]} K.G. Wilson, {\it Rev. Mod. Phys.} {\bf 47}
(1975), 773; For an exposition very close to the steps taken
here see C. Itzykson and J. -M. Drouffe, {\it Field theory},
Chapter five, Cambridge University Press, 1989.

\item{[14]} K.S. Stelle, {\it Phys. Rev.} {\bf D16}, 953
(1977).

\item{[15]} E. Fradkin and A. Tseytlin, {\it Nucl. Phys.}

\item{[16]} E. Tomboulis, {\it Phys. Lett.} {\bf 70B},
361 (1977).

\item{[17]} A.H. Chamseddine, {\it Phys. Lett.} {\bf
B332}, 349 (1994).

\item{[18]} For a discussion of quadratic divergences and
the hierarchy problem in the standard model see e.g. J.
Ellis, ``Supersymmetry and Grand Unification'', hep-ph/9512335.

\item{[19]} For a review see G. Ross, Grand unified
theories, {\it Frontiers in Physics Series}, Vol.60
(Benjamin, New York 1985).

\item{[20]} E. Witten, ``Strong Coupling Expansion of Calabi-Yau
Compactification'', hep-th/9602070.

\item{[21]} M. B\'eg, C. Panagiotakopoulos and A. Sirlin,
{\it Phys. Rev. Lett.} {\bf 52}, 883 (1984); M. Lindner,
{\it Z. Phys.} {\bf C31}, 295 (1986).

\item{[22]} For an extensive review see M. Sher, {\it
Phys. Rep.} {\bf 179}, 273 (1989).

\bye